# Phenomenological order parameter and local parameters fluctuation far beyond the critical region of the continuous phase transition

## Vladimir Dmitriev

SNBL at ESRF, CS40220, 38043 Grenoble, France; and Southern Federal University, 344092 Rostov-on-Don, Russia

E-mail: dmitriev@esrf.fr

## Abstract

In the framework of an extended phenomenological approach to phase transitions, it is shown that existing nonlinear relation between local critical atomic parameters and phenomenological order parameter induces the corresponding nonlinear temperature scaling transformation. An explicit form of such a transformation was found. Theoretically predicted uniform function reproduces well the experimentally observed behavior of order parameters in different systems.

Keywords: phase transitions; phenomenological theory; order parameter; non-critical fluctuations; scaling transformation

# 1  Introduction

It seems to be well established in the literature that despite the fact that the phenomenological Landau theory of phase transitions [1,2] deals with various types of variational free energy it predicts in fact a single set of the critical point exponents ($\alpha=0$, $\beta=1/2$, $\gamma=1$ and $\delta=3$) and thus belongs to the single "mean field" universal class [3]. We use here and in what follows the terminology of the theory of *critical phenomena* considering conveniently their terms as suitable also for a *phenomenological theory*, although this latter does not treat critical phenomena properly. The phenomenological theory characterizes a system by the same critical exponents both outside and within the critical region in which temperature behavior of a general function *f(t)* can be approximated by simple power function $f(t) = At^\lambda$, where $\lambda$ is a critical point exponent, and $t = |T - T_c|/T_c$ serves as a dimensionless variable to measure the difference in temperature from the critical temperature $T_c$. A considerable body of experimental data accumulated over the years indicates that the real systems show regular deviation from the behavior predicted in the framework of the Landau phenomenological theory and different universality classes were found experimentally in such systems. It is convenient currently to consider such a discrepancy as caused by two main reasons: (i) the phenomenological theory neglects critical fluctuations, i.e. one assumes that the order parameter can be characterized by a single value at any temperature, and (ii) the above theory uses the dimensionality for the fluctuation space lower than the marginal dimensionality ($d<d^*$) corresponding to the case (see, for example, [4-6]). One can see that the reasons both relate to the *critical* fluctuations and are valid, therefore, in the critical region. In this paper we will show that, in addition to just mentioned, another cause can lead to divergence of phenomenological calculations and the corresponding experimental data. A transcendental relation between a phenomenological order parameter and the corresponding local atomic variables as well as a nonlinear temperature scaling transformation, accompanying a change of variables just mentioned, result in a deviation of the values experimentally found for the critical exponents from those predicted in the framework of the Landau theory. Parameters of such a nonlinear transformation will be shown to depend on fluctuation properties of real systems *far beyond* the critical region.

# 2  Microscopic and macroscopic order parameters

*2.1  Order parameter space*

As a starting point of our consideration let us analyze the primary hypothesis of the Landau approach to phase transitions. The Landau theory uses the increment $\delta\rho(\mathbf{r})$ of the probability density, expressing the difference between the initial density in high-symmetry parent phase, $\rho_0(\mathbf{r})$, and the final low-symmetry phase, $\rho_d(\mathbf{r})$, expanded as a function of the basis functions of an irreducible representation (IR) $\tau_{kj}$ of the space group $G_0$ of the parent phase [1,2]. This expansion has the form:

$$\delta\rho(r) = \rho_d(r) - \rho_0(r) = \sum_{k,j,i} \eta_{kj}^i \varphi_{kj}^i(r). \quad (1)$$

The wave vector **k**, located in the first Brillouin zone, characterizes the translational symmetry of the basis functions $\varphi_{kj}^i(r)$, which are *linear* combinations of the local atomic functions, associated with the crystalline structure. The index $j$ labels the representations $\tau_{kj}$ of $G_0$, and the index $i$ ($i=1,..., n$) runs over the distinct basis functions spanning the *n*-dimensional IR $\tau_{kj}$. For a given $j$, the set of scalar coefficients $\eta_{kj}^i$ defines the *order parameter* (OP), which describes the total distortion of the initial structure at the transition. Due to the fact that usually a single irreducible OP is symmetry breaking in a phase transition, so far only necessary running index $i$ will be kept in Eq. (1).

The linear coupling between $\eta_i$ and $\varphi_i(\mathbf{r})$ in Eq. (1) means that either of these two quantities can be chosen as forming the basis of the relevant IR. As a consequence, the non-equilibrium thermodynamic potential associated with the transition, $\Phi_L(T,p,\delta\rho)$, can be considered as a function of the $\eta_i$ instead of the $\varphi_i(\mathbf{r})$. The OP components define the order parameter space $\varepsilon_n$ that is irreducible invariant space by the group $G_0$. The $\delta\rho$ variation of the probability density associated with a phase transition can be considered as a vector in the representation $\varepsilon_n$-space, and the components of $\eta = \{\eta_i^{eq}\}$ invariant vector, in the basis of this space, are the values of the OP which minimize the thermodynamic potential (for details see [7-9] and references therein).

It seems quite natural, in the absence of indications to the contrary, to consider the symmetry identity of $\eta_i$ and $\varphi_i(\mathbf{r})$ as a general property related also to their *functional form*, i.e. these values are the identical functions of the thermodynamic variables. Moreover, such an assumption seemed to be legitimate even in the framework of more general renormalization group approach. A linear *projection operator* of a space group induces basis functions for the relevant IR in the form of linear combinations of local atomic functions. These latter are characteristic of the phase transition mechanism and, in turn, were selected as a result of a *regular* renormalization transformation separating critical and non-critical variables [6,10,11]. The integral over the non-critical variables then results in the equilibrium part of the free energy $\Phi_0$ and unintegrated part forms the variational free energy (Landau potential) $\Phi_L(\delta\rho)$.

The analysis of crystal geometry of different displacive type structural phase transitions has already shown that there exists a transcendental functional relation between the value of the phenomenological OP $\eta_i$ and the magnitude of local atomic shifts [8,12,13]. The same type of non-linear periodic dependence has been obtained as well for $\eta_i$ as a function of probability density variation for the segregation type phase transitions [14]. Thus, analyzing the transformation mechanisms, one can conclude that the order-parameter space, denoted hereafter $\sigma_n$, in general case (i.e. for the full range of the OP variation) differs in an essential manner from the order-parameter space $\varepsilon_n$ used earlier in the description of continuous phase transitions. While $\varepsilon_n$ is a *n*-dimensional *vectorial* space, $\sigma_n$ is a *n*-dimensional closed *functional* space *with boundary*, the structure of which depends on the type of variational parameters identified the transition mechanism [9].

*2.2 Phenomenological order parameter and essential variational parameters*

In order to derive the general form of the function $\eta_i(\xi_j)$, where the set of $\eta_i$ is a long-range phenomenological order parameter and $\xi_j$ represent variational local atomic parameters, i.e. short-range order parameters (variation of probability for the segregation or disorder-order transformation, magnitude of atomic displacements for displacive type transitions etc.) let us consider the problem in the functional order-parameter space $\sigma_n$.

One can make use of the usual approximate scheme for calculating $\eta(\xi)$ by finding that solution of the Euler's variational equation $\{\delta\Phi/\delta\eta(\xi)\}=0$ which minimizes the free energy functional. The appropriate choice for the latter in the case of continuous phase transition is the classical Landau-Ginzburg functional

$$\Phi(\eta) = \int_{V_\xi} d\xi \left\{ a_1 \eta(\xi)^2 + a_2 \eta(\xi)^4 + g \left( \frac{d\eta}{d\xi} \right)^2 \right\}, \qquad (2)$$

where the integral is over a volume in the *OP space*. For the sake of simplicity we treat a single-component or effective OP and the conjugate external field is neglected. The coefficient $a_1$ is conveniently assumed to be a regular function of thermodynamic variables (the temperature, pressure, etc.) whereas the remaining coefficients are regarded as temperature and pressure-independent parameters. The corresponding Euler's equation has form:

$$\frac{\delta\Phi}{\delta\eta(\xi)} = a_1 \eta(\xi) + 2a_2 \eta(\xi)^3 + g \frac{d^2\eta}{d\xi^2} = 0. \qquad (3)$$

The boundary conditions are $\eta(0)=0$ and $\eta'(0)=1$. The first condition indicates a coincidence of the origin points for the variables $\eta$ and $\xi$. The second one ensures their identical behavior close to $T_c$, i.e. it justifies the change of variables $\xi \to \eta$ in the Landau theory. The differential equation (3) has exact general solution which is expressed as

$$\eta = \eta_0 sn[\mu(\xi - \xi_0), \kappa]. \tag{4}$$

In the above equation $sn[\mu(\xi - \xi_0), \kappa] = snu$ is the elliptic sine of Jacobi, $\mu$ and $\xi_0$ are arbitrary constants [15]. Taking into account the boundary conditions one gets $\mu=1$ and $\xi_0=0$. The parameter $\kappa = (a_2/|g|)^{1/2}$ is the *modulus* of the elliptic integral of the first kind

$$u = F(\xi, \kappa) = \int_0^\xi \frac{d\xi}{\sqrt{1 - \kappa^2 \sin^2 \xi}}. \tag{5}$$

The phase diagram of the considered system can be worked out by minimizing the Landau potential

$$\Phi_L(\eta) = a_1 \eta^2 + a_2 \eta^4, \tag{6}$$

where the OP $\eta$ has the form of Eq. (4). The minimization of $\Phi_L$ with respect to the actual variational parameter $\xi$ is expressed by $\frac{\partial \Phi_L}{\partial \xi} = \frac{\partial \Phi_L}{\partial \eta} \cdot \frac{\partial \eta}{\partial \xi}$ which gives the equation of state

$$\eta \eta' \{a_1 + 2a_2 \eta^2\} = snu \cdot cnu \cdot dnu \{a_1 + 2a_2 \eta_0^2 sn^2 u\} = 0, \tag{7}$$

where $cnu = \sqrt{1 - sn^2 u}$ and $dnu = \sqrt{1 - \kappa^2 \sin^2 \xi}$.

Equation (7) yields three possible stable states: (i) The parent phase *I* for $snu=0$ (origin of the space σ); (ii) The limit, non-Landau, phase *II*, given by $cnu=0$ ($snu=1$) (boundary of the space σ), corresponding to the fixed values $\eta_0$ of the OP; (iii) "Landau " phase corresponding to the standard minimization of $\Phi_L$ with respect to the OP $\eta$, whose value $\eta^2 = -a_1/2a_2$ varies between 0 and $\eta_0$ (interior of the space σ). The function *dnu* has zeros only if κ=1, however even in this case *dnu* vanishes simultaneously with *cnu* and no different solution of Eq. (7) exists.

Then the stability condition has the form:

$$\frac{\partial^2 \Phi_L}{\partial \eta^2}\left(\frac{\partial \eta}{\partial \xi}\right)^2 + \frac{\partial \Phi_L}{\partial \eta}\frac{\partial^2 \eta}{\partial \xi^2} \geq 0. \tag{8}$$

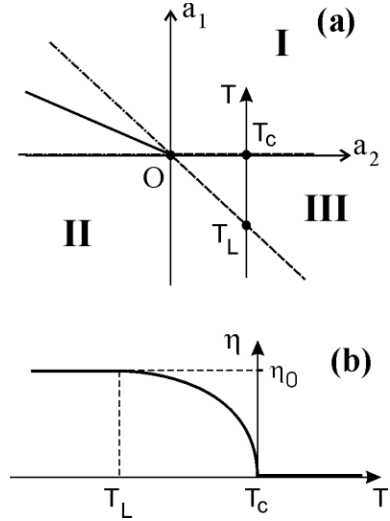

**Figure 1** (a) Phase diagram of the model Eq. (6). Solid, dashed and dashed-dotted lines are respectively discontinuous first-order transition, continuous transition, and limit of stability lines. *O* is the three-phase point. (b) Temperature dependence of the order parameter for the path indicated in (a).

The corresponding phase diagram in the plane of thermodynamic parameters ($a_1, a_2$) is shown in Fig. 1(a). The second-order phase transition line $a_1 = 0$ separates parent *I* and Landau *III* phases. The stability regions of Landau *III* and limit *II* phases adjoin along the line $a_1 = -2a_2$. The first-order transitions line $a_1 = -a_2$ between parent *I* and limit *II* phases meet, in the three-phase point *O*, two latter lines. It is worthwhile to note that in the classical Landau theory the obligatory positivity of the coefficient $a_2$ in the potential (6) ensures the global stability of the phases, and the convexity of the potential for large values of the OP. However, equation (4) shows that magnitude of the OP cannot be arbitrary large. This restriction raises the possibility of global stable phase diagrams, even for negative values of $a_2$ [Fig. 1(a)] [9].

The important conclusion stems from the above consideration. One can see in Fig. 1(a) that the temperature range of the OP variation, that is, the stability region of Landau phase *III*, is limited. Let us denote by $T_L$ the low-temperature stability limit for the Landau phase [Fig. 1(a)]. The order parameter has value 0 above $T_c$, varies between 0 and $\eta_0$ for $T_c > T > T_L$, reaches this saturation value $\eta_0$ (conventionally, $\eta_0 = 1$) at $T_L$ and then remains invariable for $T < T_L$ [Fig.1(b)]. Since there are no regular reasons for $T_L = 0$ K, one should make use of a generalized form for the dimensionless variable $t_n = (T - T_c)/(T_c - T_L)$. The latter distinguishes the Landau approach, in which any phase has its thermodynamic limit of stability at a finite temperature (or pressure) $T_L \neq 0$ K defined by Eq. (8) $\partial^2 \Phi_L / \partial \xi^2 = 0$ where the corresponding energy minima disappear, and the Gibbs type consideration in which the set of minima exists at any temperature up to $T = 0$ K but the system finds a lowest one for any given *T*.

## 3 Temperature scaling transformation

### 3.1 OP temperature behavior

We already discussed in Sec. 2.1 that it is convenient in the Landau theory to use identical temperature dependencies for both a phenomenological OP $\eta$ and atomic variable parameters $\xi$. This means that the Landau potential identical with Eq. (6) can be introduced for these latter as well: $\tilde{\Phi}_L(\xi) = \alpha_1 \xi^2 + \alpha_2 \xi^4$. By minimizing variational free energy $\tilde{\Phi}_L$ and assuming as usual $\alpha_1 = \alpha_{10} t$ ($\alpha_{10}$=const) one can find $\xi = (-\alpha_1 / 2\alpha_2)^{1/2} \propto t^{1/2}$. The consideration in Sec. 2.2 has shown also the existence of the complex relation between essential variational parameters $\xi_i$ and measurable phenomenological OP $\eta_j$. Thus, if one approximates $\xi = \tilde{\alpha}_{10} t_n^{1/2}$, then a phenomenological OP $\eta$ in function of temperature, following Eq. (4), will have form

$$\eta(t_n) = \eta_0 sn(\tilde{\alpha}_{10} t_n^{1/2}; \kappa). \tag{9}$$

Elliptic sine is considered by definition as a circular sine whose argument is scaled with the integral transformation (5): $sn(x,\kappa)=sin\,\tilde{x}$ (factor $\pi/2$ is included in the definition of the argument) [15]. Making use of temperature as a variational thermodynamic parameter one can convert Eq. (5) into nonlinear temperature scaling transformation

$$\sqrt{\tilde{t}} = \int_0^{t_n} \frac{d(\sqrt{t_n})}{\sqrt{1-\kappa^2 sin^2(\frac{\pi}{2}\sqrt{t_n})}}. \tag{10}$$

Nonlinearity of such a temperature scaling transformation is fully specified by the modulus $\kappa$ varying from 0 to 1. Figure 2 shows scaling ratios between variables $t_n$ and $\tilde{t}$ for different values of the modulus $\kappa$. The simplest situation is that $\kappa=0$ and thus the transformation is linear. It is worth reminding, that the corresponding sinusoidal form for OP as a function of $\xi$ was already used earlier in the phenomenological models of displacive type reconstructive phase transitions in crystals [9,12,13] and of segregation transformations in complex fluids [14]. However, the above consideration complements the approach with temperature-controlled fluctuations.

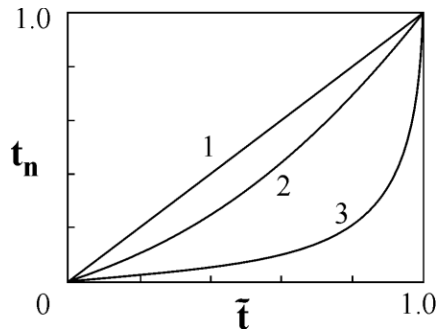

**Figure 2** Scaling ratio between temperature variables $t_n$ and $\tilde{t}$ for different values of the modulus $\kappa$: 1-0, 2-0.8, 3-0.999.

In order to confirm an applicability of the present approach to real physical systems we have chosen three crystals undergoing continuous phase transitions of different nature. The choice was dictated by the existence for the crystals of reliable experimental data approved with different techniques available.

1) In metallic *Sn* transition to the *superconducting* state takes place at $T_c=3.782$ K [16]. While the microscopic OP for such a transition was chosen as an effective wave function $\Psi(\mathbf{r})$ which, in turn, was shown to be proportional to the local value of the energy-gap parameter $\Delta$ [17,18], the measurable phenomenological OP is average normalized energy gap in the elementary excitation spectrum of superconductor [16].

2) The *cooperative Jahn-Teller* transition where the ordering occurs in the electronic degrees of freedom reduces symmetry of the perovskite $PrAlO_3$ crystal from tetragonal to monoclinic at $T_c=151$ K. The electronic functions of the lowest crystal-field double degenerated level $E_g$ of the $Pr^{3+}$ (in the parent cubic field) form the basis functions of the relevant IR. The observable phenomenological OP can be chosen either as the splitting between the doublet levels in the ordered phase or as linearly coupled acoustic- and optical-phonon modes of the same symmetry [19].

3) *Disorder to order* transformation in β-brass was found to occur at $T_c=740$ K [20-22]. It is convenient to consider the deviation of average probability of atoms to occupy positions in a crystal lattice as the phenomenological OP for ordering type transitions. The local variation of the probability of occupation or, equivalently, population of the corresponding sites is considered to be essential variable parameter for the transition.

The temperature $T_L$, limiting stability region of Landau phase, and the modulus $\kappa$, characterizing nonlinearity of the temperature scaling transformation were used as the fitting parameters. Figure 3 shows the best fit for different OPs plotted versus *scaled normalized* temperature $\tilde{t}$ and compared with the worked out function $\eta(\tilde{t}) = sin\frac{\pi}{2}\sqrt{\tilde{t}}$. Only temperature range of interest in which $\eta \neq$const is displayed. The corresponding fitting parameters were fixed as follows: *Sn* ($T_L=1.236$ K; $\kappa=0.05$), $PrAlO_3$ ($T_L=61$ K; $\kappa=0$) and β-brass ($T_L=490$ K; $\kappa=0.77$). One can see that the uniform function predicted in the framework of the present approach reproduces fairly well the experimentally observed behavior of OPs in different systems. To show clearly the role of the temperature scaling transformation, experimental spots for the β-brass OP [21,23] are plotted, in the inset on Fig. 3, also versus the *non-scaled* normalized temperature $t_n$. Two curves in the inset display functions $\eta_1 = sin\frac{\pi}{2}t_n^{1/2}$ (solid line) and $\eta_2 = sin\frac{\pi}{2}t_n^{1/3}$ (dashed line). It is clear that close to $T_c$ the approximate function $\eta_2 \propto t_n^{1/3}$ reproduces more sufficiently experimentally observed OP behavior, and one could deduce for β-brass the critical exponent β≈1/3.

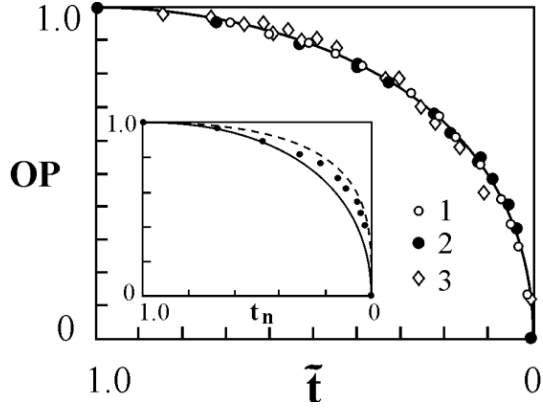

**Figure 3** Reduced values of the order parameters for (1) *Sn* [16], (2) β-brass [22], and (3) *PrAlO$_3$*[19] as a function of rescaled temperature, and compared to the theoretical curve (solid line). For the inset see the text.

### *3.2 Modulus of the temperature scaling transformation*

In view of the preceding consideration one can conclude that the phenomenological approach deals with two temperature scales. The first one can be termed *real-temperature*. This temperature *t* (or, equivalently, $t_n$) is the *measurable* thermodynamic variable for the observable functions describing behavior of a system. Another, *rescaled* temperature $\tilde{t}$ appears in the phenomenological approach after a renormalization procedure eliminated local microscopic variables and replaced them by the macroscopic averaged variables those are components of the phenomenological OP. Precisely, they are the rescaled variables the Landau theory deals with considering temperature behavior of different physical quantities or obtaining the corresponding critical exponents. The example of β-brass (see Fig. 3) shows clearly that different value for the OP critical exponent can be obtained depending on the temperature scale used. To compare experimental results and phenomenological predictions one should take into account the temperature scaling transformation Eq. (10). Figure 4 shows the real-temperature OP critical exponent *β* in function of the modulus *κ* of such a transformation. Thus, despite of the fact that the value *β*=1/2 is predicted unambiguously in the framework of the Landau theory it would be expectable to get *different* values for the critical exponent measured in the real-temperature scale.

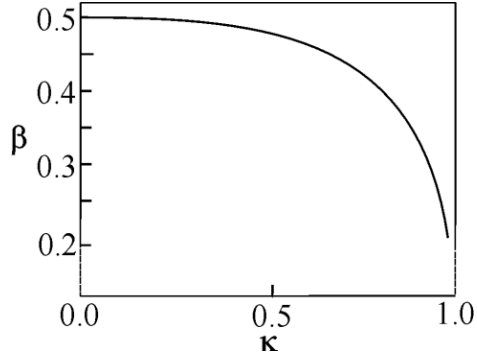

**Figure 4** Measurable real temperature order parameter critical exponent as function of the transformation modulus κ.

Figure 5 summarizes transformation steps, which an experimentally measured curve $\eta(t)$ should pass to be compared with predictions of a phenomenological theory. Customary normalization of the temperature scale by the corresponding critical temperature $T_c$ brings different curves to the single origin [Fig. 5(a)]. Then modified normalization introduced in Sec. 2.2 merges curves as at the origin $t_n=0$ ($T=T_c$) as at the saturation point $t_n=1$ ($T=T_L$) of the OP variation range [Fig. 5(b)]. Finally, nonlinear temperature scaling transformation $t_n \to \tilde{t}$ fits experimental points for different systems in the uniform curve $\eta(\tilde{t}) = sin\frac{\pi}{2}\sqrt{\tilde{t}}$ [Fig. 5(c)], predicted in the framework of an extended phenomenological approach.

In order to elucidate the physical meaning of the modulus $\kappa$ of the nonlinear transformation (10) let us consider the response function $\tilde{G}(\xi) = (4\pi g\xi)^{-1} e^{-q\xi}$, where $q^{-2} = g\chi$, and $\chi$ is the generalized susceptibility [23]. The quantity $r_c = q^{-1}$ is the correlation length of the short-range order. The latter can be directly measured in a scattering experiment when the radiation couples to the OP [5]. By incorporating a generalized external field in Eq. (6) it is easy to calculate the function $\chi = (a_1 + 3a_2\eta^2)^{-1}$ and then derive for the modulus $\kappa^2 = a_2/|g| = \varepsilon r_{c0}^{-2}$, where $r_{c0}$ is the correlation length far from the phase transition point ($t\sim1$) and ε is the normalizing factor. Thus, one can conclude that the modulus κ is characteristic of the *background* (i.e. far beyond the critical region) fluctuational property of a physical system.

*3.3 Landau-to-limit phase transition*
We next discuss the specific features of the non-Landau limit phase and manifestation of the Landau-to-limit-state transformation. The limit phase was typified in the above consideration (Sec. 2.2), with respect to the classical Landau phase, by the property of the OP to retain maximal value $\eta_0$=const independent of temperature variation. A considerable body of experimental data indicates clearly that in different physical systems such states do exist. The substances considered in Sec. 3.1 provide typical examples of the temperature behavior predicted in the framework of the present approach [see Fig. 1(b)]. It should be stressed that in the limit state not necessarily perfect but maximally possible, for the given system, degree of order $\eta_0 \leq 1$ can be realized. This means that in the present approach the limit phase is the state with temperature independent stable equilibrium between perfect order and partial disorder. One deals with a *saturated* state and perfectly ordered one is a particular case. The latter property along with the absence at $T_L$ of a singularity in the function $\eta(t)$ provides specific appearance of the Landau-to-limit-state transition. One can show that such measurable thermodynamic functions as entropy, latent heat or specific heat, are proportional to the derivative $\frac{\partial \eta}{\partial t} = \eta_0 \cos \frac{\pi}{2}\tilde{t}$. However, this latter has no singularities and it is equal to zero at $T_L$ (or, equivalently, at $\tilde{t}=1$). Thus, they show no jump-like behavior, divergence or different singularities at the corresponding transition point. From this, it is clear that the Landau-to-limit-state transformation cannot be identified neither as first nor as second order phase transition. More detailed analysis of the features of such a high order transition is beyond the scope of this paper and will be subject of a forthcoming report.

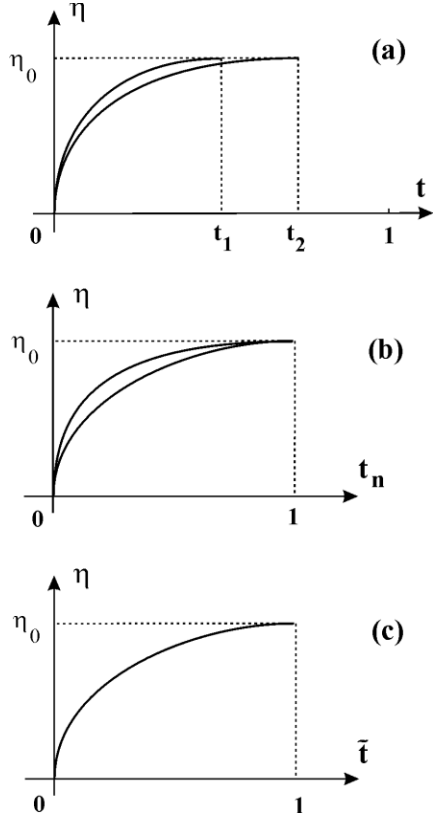

**Figure 5** Order parameter variation curves presented in different temperature scales. (a) OP as a function of a dimensionless variable $t=|T-T_C|/T_C$, (b) as a function of $t_n=(T-T_C)/(T_C-T_L)$, and (c) in function of the nonlinearly rescaled variable $\tilde{t}$.

## 4 Conclusions and Outlooks

Two assumptions were involved in the above consideration and should be discussed to ensure generality of the present approach. First, only the case of the single dimensional OP was treated in this work. However, temperature behavior of the OP, in the Landau type phenomenological approach, is controlled by the coefficient $a_1$ [see Eqs. (2) and (6)] associated with the invariant quadratic in the OP components $a_1 I_1 = a_1(\eta_1^2 + ... + \eta_n^2)$ of the free energy expansion, and exactly $a_1$ specifies therefore an uniform functional form of the temperature dependence $\eta_i(t)$ for each of multiple OP components as well as for the corresponding single component effective OP. One can find a further confirmation of such a conclusion in Fig. 3 where the uniform function reproduces well temperature behavior of the single- (β-brass), two- ($Sn$) and three-component ($PrAlO_3$) phenomenological OPs.

Another point is that the present consideration deals with the free energy expansion restricted to the fourth degree term in OP components. However, again, the argument concerning uniqueness of the coefficient $a_1$, which specifies the OP temperature behavior, is valid. Incorporation of higher order terms modifies stability conditions and form of stability boundaries and transition lines in the phase diagram but does not affect temperature-dependent part of the free energy.

Summarizing, we have shown in the framework of an extended phenomenological approach that there exists a uniform function describing the OP temperature variation. The renormalization procedure eliminated local atomic variables and replaced them by the phenomenological order parameter components is accompanied by the nonlinear temperature scaling transformation. *Rescaled* temperature is an essential thermodynamic variational parameter of the Landau theory with respect to which mean-field critical exponents are predicted unambiguously while the corresponding experimental values are obtained in the *real-temperature* scale. The general form of such a transformation is found to be elliptic integral of the first kind.

## 5 Acknowledgements

Drs I Lebedyuk and A Dmitriev are warmly acknowledged for their assistance in computer calculations.